\documentclass[sigconf]{acmart}

\usepackage[group-separator={,},group-minimum-digits=4]{siunitx}
\usepackage[inline]{enumitem}
\usepackage{booktabs}
\usepackage{listings}
\usepackage[capitalise]{cleveref}
\usepackage{lipsum}
\usepackage{timing-diagrams}
\usepackage{url}
\usepackage{multirow}
\usepackage{csquotes}
\usepackage{multicol}

\acmConference[MSR 2022]{MSR '22: Proceedings of the 19th International Conference on Mining Software Repositories}{May 23-24, 2022}{Pittsburgh, PA, USA}

\newcommand{\api}{\textsc{API}}
\newcommand{\cor}{code review}
\newcommand{\diff}{differential revision}
\newcommand{\ger}{{G}errit}
\newcommand{\gh}{{G}it{H}ub}
\newcommand{\json}{\textsc{JSON}}
\newcommand{\oss}{open-source software}
\newcommand{\phab}{{P}habricator}

\begin{document}

\title{The Unexplored Treasure Trove of Phabricator Code Reviews}

\begin{abstract}
\phab\ is a modern code collaboration tool used by popular projects like {F}ree{BSD} and {M}ozilla. 
However, unlike the other well-known code review environments, such as
\ger\ or \gh, there is no readily accessible public \cor\ dataset for \phab.
This paper describes our experience mining \cor s from five different
projects that use \phab\ ({B}lender, {FreeBSD}, \textsc{KDE},
\textsc{LLVM}, and {M}ozilla).
We discuss the challenges associated with the data retrieval process
and our solutions, resulting in a dataset with details regarding 
\num{317476} \phab\ \cor s.
Our dataset\footnote{\url{{https://doi.org/10.6084/m9.figshare.17139245}}}
is available in both \json\ and {MySQL}
database dump formats.
The dataset enables analyses of the history of \cor s
at a more granular level than other platforms.
In addition, given that the projects we mined are publicly accessible via
the {C}onduit \api~\cite{conduit_api_overview},
our dataset can be used as a foundation to fetch
additional details and insights.
\end{abstract}

\author[1]{Gunnar Kudrjavets}
\orcid{0000-0003-3730-4692}
\affiliation{
   \institution{University of Groningen}
   \city{Groningen}
   \country{Netherlands}}
\email{g.kudrjavets@rug.nl}

\author[2]{Nachiappan Nagappan}
\orcid{0000-0003-1358-4124}
\affiliation{
    \institution{Microsoft Research}
    \streetaddress{One Microsoft Way}
    \city{Redmond}
    \state{WA}
    \country{USA}
    \postcode{98052}}
\email{nnagappan@acm.org}
\authornote{This work was initiated while Nachiappan Nagappan was with Microsoft Research. He is currently with Meta Platforms, Inc.}

\author[3]{Ayushi Rastogi}
\orcid{0000-0002-0939-6887}
\affiliation{
   \institution{University of Groningen}
   \city{Groningen}
   \country{Netherlands}}
\email{a.rastogi@rug.nl}



\maketitle

\section{Introduction}

A variety of \cor\ datasets are published.
Some of the most well-known include
{C}ode {R}eview {O}pen {P}latform (\textsc{CROP})~\cite{crop_dataset},
{G}errit \cor\ dataset~\cite{gerrit_dataset}, and
{GHT}orrent~\cite{ghtorrent_dataset}.
Several popular \oss\ projects (e.g., \textsc{F}ree{BSD}, \textsc{LLVM}, {M}ozilla)
use a code collaboration tool called \phab~\cite{phacility} to conduct their \cor s.
We have not found any published \cor\ datasets for \phab.
The search of existing literature about mining popular code collaboration tools
reveals a study documenting the mining of
{G}errit data for {A}ndroid~\cite{mukadam_2013} and \gh~\cite{ghtorrent_dataset}.
We can locate only one thesis about mining projects using \phab~\cite{cotet_2019}.
This thesis describes the development of a data mining tool called Phabry~\cite{phabry_2021}.
Phabry, however, cannot be used to collect code changes associated with a \cor.

The absence of a readily accessible dataset of code changes for \phab\ projects
has deprived \cor\ researchers of a rich information source.
The benefit of \phab\ is the ability to
formally distinguish between different events taking place during
the \cor.
Each event and action taken during the life cycle of a \cor\ is associated with an
author's identity and an event's timestamp.
Researchers can track when a \cor\ was accepted, abandoned,
taken over by someone else, when a reviewer resigned,
when some attributes (e.g., title) were updated, etc.

\begin{table}[!htbp]
\centering
\caption{Different events during \phab\ \cor.}
\label{tab:phab-events}
\begin{tabular}{llll}
\toprule
abandon     & create                & reopen                &  subscribers\\
accept      & draft                 & request-changes       &  summary \\
author      & inline                & request-review        &  testPlan\\
close       & plan-changes          & resign                &  title \\
commandeer  & projects              & reviewers             & update\\
comment     & reclaim               & status                & \\
\bottomrule
\end{tabular}
\end{table}

\Cref{tab:phab-events} lists all possible events we noted during the
\cor\ life cycle of popular \phab\ projects.
We are not aware of any other \cor\ system that tracks events with
this level of granularity.
{G}it{H}ub introduced the more formalized \cor\ process, including
functionality for actions such as formal acceptance of code changes,
only in 2016~\cite{github_code_review_2016}.
By not utilizing publicly available \phab\ data,
researchers miss out on potentially valuable
insights and opportunities to study influential and popular software projects
with a multi-year development history.

Without a pre-existing accessible dataset, we set out to acquire the
\phab\ data ourselves and convert the data to a format suitable for
further analysis.
Based on our experience, reliably mining data associated with hundreds of
thousands of \cor s,
even with a pre-existing tool, is an involved and time-consuming
process requiring a nontrivial amount of manual labor.
We describe the challenges encountered and our solutions in~\Cref{subsec:challenges}.

The primary motivation behind our paper is to publish a dataset that
\begin{enumerate*}[label=(\alph*),before=\unskip{ }, itemjoin={{, }}, itemjoin*={{, and }}]
    \item does not require extra mining effort
    \item includes data about code changes in the \cor s (files changed; lines of code added, deleted, or updated)
    \item can be imported into a relational database system such as {M}y{SQL} in addition to being published in a plain \json\ format.
\end{enumerate*}

\section{History and overview}

\phab\ was initially developed as an internal \cor\ tool for Facebook
in 2011~\cite{tsotsis_meet_phabrictor}.
As of this paper (November 2021), it is still the de facto \cor\ environment
for Facebook and is internally under active development.
The public version of \phab\ is developed by a company called Phacility
and distributed as \oss~\cite{phacility}.

When compared to other well-known \cor\ environments, such as \ger\ or \gh,
\phab\ introduces some new \cor\ related terminology.
For example, the proposed code modifications in \ger\ are referred to as \emph{change} (same as \emph{pull requests} in the context of \gh).
A \cor\ iteration in {G}errit is a version of the change and is called \emph{patch set}.
In \phab\ both the initial set of code modifications and its subsequent versions are called 
\emph{differential revision}, which gets shortened to a \emph{diff}.
Committing and merging the accepted changes to the target branch is called
\emph{submitting} in \ger\ and \emph{landing} in \phab.

\begin{table}[!htbp]
\centering
\caption{Descriptive data about \phab\ projects.}
\label{tab:phab-projects-overview}
\begin{tabular}{llcrr}
\toprule
\multicolumn{1}{c}{Name} &
	\multicolumn{1}{c}{Type of} &
		\multicolumn{1}{c}{Year of} &
			\multicolumn{1}{c}{Total} &
				\multicolumn{1}{c}{Accessible} \\
& \multicolumn{1}{c}{software}&
	\multicolumn{1}{c}{first diff} &
		\multicolumn{1}{c}{reviews} &
			\multicolumn{1}{c}{reviews} \\
\midrule
Blender~\cite{blender_phab} & Graphics & 2013 & \num{13151} & \num{13097} (99.59\%) \\
{F}ree{BSD}~\cite{freebsd_phab} & \textsc{OS} & 2013 & \num{32884} & \num{32725} (99.52\%) \\
\textsc{KDE}~\cite{kde_phab} & Desktop & 2015 & \num{29953} & \num{29874} (99.73\%) \\
\textsc{LLVM}~\cite{llvm_phab} & Compiler & 2012 & \num{113372} & \num{112892} (99.58\%) \\
{M}ozilla~\cite{mozilla_phab} & Browser & 2017 & \num{130567} & \num{128888} (98.71\%) \\
\bottomrule
\end{tabular}
\end{table}

A wide range of \phab\ projects are publicly accessible.
\Cref{tab:phab-projects-overview} lists the projects published by our dataset.
We describe the type of project,
when the first \diff\ was published,
the amount of available \cor\ data as of November 2021,
and the percentage of \cor s that are publicly accessible.
The median age of a project is 8 years and the
median number of unique contributors per project is \num{1504}.
Out of \num{317476} \cor s, only \num{258} (0.08\%) do not have any associated
data quantifying the code changes.
The lack of data is caused by changes consisting of binary files or containing only renaming of files.

\section{Mining data}

\subsection{Authentication and data access}

The \phab\ user community maintains a list of organizations and projects
that utilize the tool~\cite{phabricator_usage}.
We find that resource, in addition to our knowledge from industry experience, to be the
best available reference related to \phab 's usage.
Interaction with \phab\ is conducted via {C}onduit \api~\cite{conduit_api_overview}.
{C}onduit \api\ is a remote procedure call protocol where requests and
responses are encoded in \json\ (\textsc{\json-RPC}).
To mine \phab\ data, the client needs to have a {C}onduit \api\ token for
authentication purposes.
Acquiring the token requires creating a user account for each \phab\
instance to be mined.
Account creation can either require a manual approval from a member of
the development team ({F}ree{BSD}),
possession of the {G}it{H}ub account ({M}ozilla, \textsc{LLVM}), or
just filling out the required registration data ({B}lender, \textsc{KDE}).

The official \api\ documentation contains only a limited number of examples about its usage.
We find that practical experimentation with \texttt{curl}~\cite{curl} or \api\ console
is the most efficient way to gain knowledge~\cite{phabricator_api}.
To mine the data, one can develop their own tool(s)
(something authors of this paper initially did) or utilize existing 
\api\ wrappers for different programming languages~\cite{conduit_tools}.
For our past \cor\ related studies, we have utilized
a version of Phabry with minor
modifications to facilitate the debugging and adjustments necessary
to mine different \phab\ instances~\cite{phabry_2021}.
We find that the thesis describing Phabry's development is a detailed and
valuable reference about how to interact with {C}onduit \api~\cite{cotet_2019}.

\subsection{Parsing and interpretation}

Data retrieved via {C}onduit \api\ is returned in \json\ format.
Our initial instinct was to follow the approach taken in both {G}errit and
{GHT}orent datasets and import the data into a database such as {M}y{SQL}~\cite{gerrit_dataset,ghtorrent_dataset}.
Though the output from {C}onduit \api\ is not documented, building a
relational normalized database schema was a straightforward process.
The downside of exposing the dataset as a database is the cost associated
with maintaining the database instance, importing data, deciding
what fields to index, etc.
For our studies, we both parse and extract data from \json\ directly and
use \textsc{SQL} to mainly gather descriptive statistics.
That approach proves to be performant even with dataset sizes between 2--3 \textsc{GB}s and up to \num{135000} files.
The dataset we expose contains both raw \json\ files and {M}y{SQL} database
containing the same information.

\subsection{Associating \diff s with code}

Each differential revision can evolve through multiple versions.
Code changes between each version can differ.
To understand the full evolution of the \cor\ it is necessary to keep
track of how the \cor\ evolved over the time.
However, most \cor\ related studies limit themselves to only
the initial or the final version of code changes.
In addition, our intent is not to duplicate the data stored in the source control system.
For our dataset, we keep track of number of files changed and
lines added, deleted, and updated for the final version of the differential
revision.
We use \texttt{diffstat} to calculate the code churn statistics from
the raw diff output~\cite{diffstat}.

There are multiple options for mapping the final code changes to \diff s.
The intuitive approach is to inspect the commit history of a source control system
and match the commit content with a \diff.
\Cref{code:freebsd-commit-sample} displays a randomly picked {F}ree{BSD}
commit using a \phab\ \cor\ process.

\lstset{
    breaklines=true,
    basicstyle=\ttfamily\scriptsize,
    keepspaces=false,
    caption={Anonymized {F}ree{BSD} commit description.},
    label=code:freebsd-commit-sample,
    frame = single
}

\begin{lstlisting}
commit mrmauqgsbpmdymqzchdtnmxmadimcakrzesmjeil
Author:     John Doe <john.doe@FreeBSD.org>
AuthorDate: 2971410770
Commit:     John Doe <john.doe@FreeBSD.org>
CommitDate: 2971410770

    foo: fix a memory corruption in bar.

    Differential Revision:  https://reviews.freebsd.org/D12345678
\end{lstlisting}

Based on our analysis, the presence of the string associating a 
commit with the specific \diff\ is \emph{optional} and depends on the project.
In addition, we observe typographic errors in the \textsc{URL}s referencing
\diff s and using different notations when referring to a \cor.
Therefore, we cannot reliably use the data from commit descriptions to determine
what \diff\ they are associated with.
Another challenge with this approach is handling the presence of many-to-many relationships~\cite{garcia-molina_database_2009}
between commits and \diff s.
A single commit can tag multiple \diff s and a single \diff\ can be
referenced from multiple commits.
Fetching the data about code changes
directly from \phab\ results in a correct representation of final code changes.

\subsection{Challenges}
\label{subsec:challenges}

\subsubsection{Networking}

The server hosting \phab\ may apply rate limiting to
the number of requests a {C}onduit \api\ client can issue or the number of
network connections the client can make overall.
We find that it is necessary to have a retry mechanism in place to mitigate
the presence of intermittent errors such
as server returning a variety of \textsc{HTTP} error codes, connections
timing out, etc.
Depending on the specifics of a \phab\ instance, the server may also
require a \textsc{HTTP}
\textsc{GET} request for one project and \textsc{PUT} request for another (e.g., Blender).

\subsubsection{Permissions}

During our data mining process we found that there is a subset of \diff s 
accessible only to authenticated users, i.e., they cannot be
directly downloaded via \texttt{curl} without providing the required {C}onduit \api\ token.
We utilize the subsequent usage of getting the metadata about \diff\ from 
\texttt{differential.query}~\cite{differential_query} and using it to fetch
the raw content by calling \texttt{differential.getrawdiff}~\cite{differential_getrawdiff}.

However, there were some revisions which even an authenticated user could not access.
Those \diff s were a minor part of the overall dataset,
accounting for a median of 0.42\% of \diff s
per \phab\ instance.

\subsubsection{\textsc{API} evolution}

\phab\ is distributed as \oss\ and each project is
free to make any changes needed for their purposes.
Depending on the \phab\ instance, the type of data returned by {C}onduit
\api\ may be different.
Differences may manifest in the data fields present,
action types that can be performed on a \diff, and
if certain fields are optional or mandatory.
In some cases, even the data type of the field varies between different \phab\ instances.

\section{Database schema}

\subsection{Design decisions and data representation}

One of the initial design decisions we faced was a choice of
exposing the data in the database as close to its original
representation in \json\ versus using a third normal form~\cite{garcia-molina_database_2009}.
Third normal form is used to reduce data duplication,
amount of storage required, and increase the performance of
database queries.
For simplicity, we chose to match the \json\ structure as
much as possible
unless normalization was needed to represent entries of
variable count.

\begin{figure*}[!htbp]
    \centering
    \includegraphics[width=2\columnwidth,keepaspectratio]{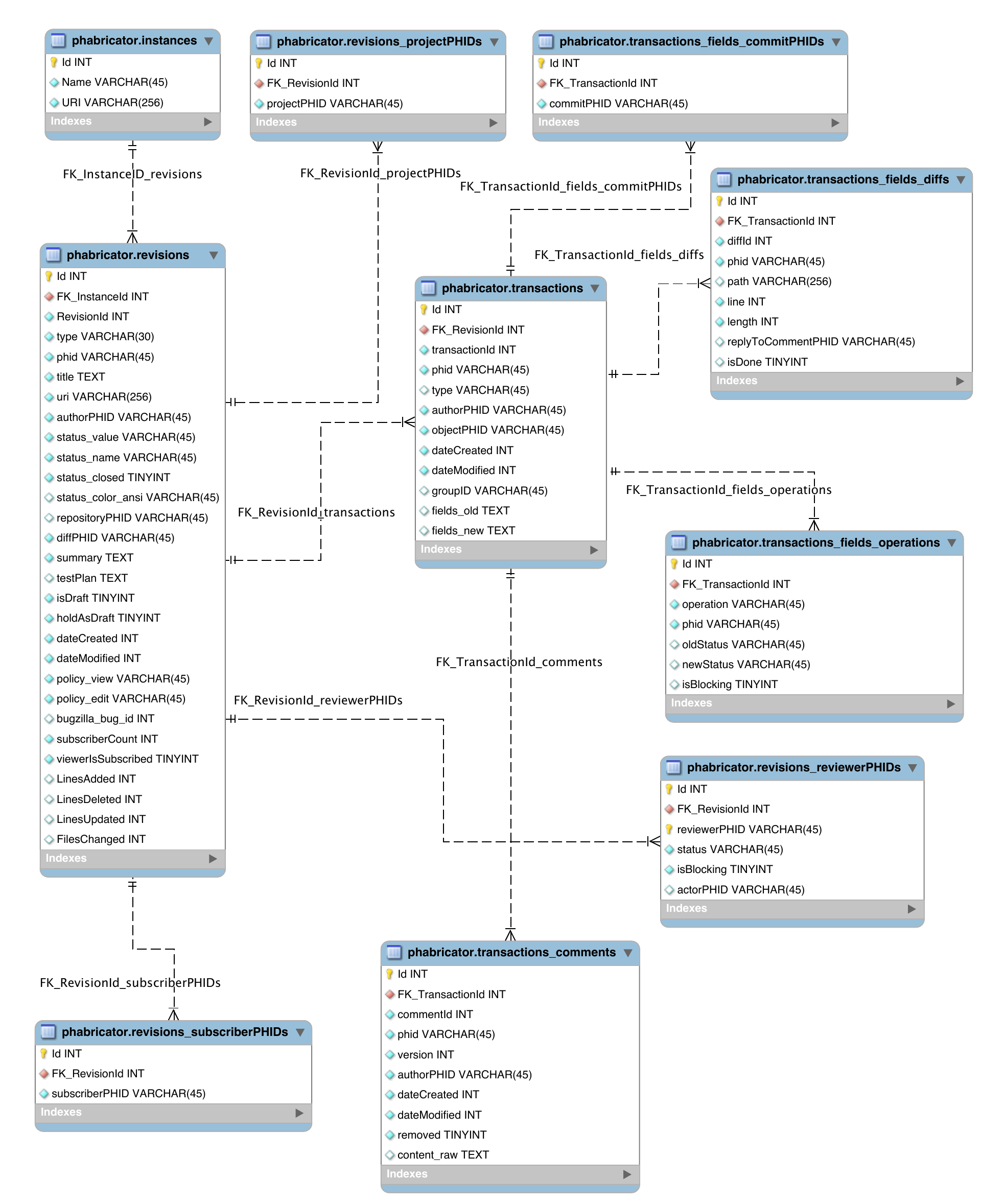}
    \caption{Database schema describing \phab\ \diff s, transactions, and associated entities.}
    \label{fig:db_schema}
\end{figure*}

The full relational database schema is presented in~\Cref{fig:db_schema}.
Each table has a primary key called \emph{Id}.
The foreign key columns referencing parent tables are prefixed with \emph{FK\_} and end with the name of a referring table.
In modeling the data, we chose to follow the Phabry output
directory structure and the \phab\ design concepts.
Two essential tables are \emph{revisions} and \emph{transactions}.
Transactions in the context of \phab\ are the history of edits associated with each
revision~\cite{transaction_search}.
Each revision belongs to a single \phab\ instance stored in the
\emph{instances} table.
One revision can have many transactions associated with it.
Each transaction belongs to only a single revision.
Each revision can have many reviewers and subscribers related to it.
A revision can belong to many projects.
Each transaction can be associated with many comments,
inline comments,
a set of changed fields, or
many commits.
The timestamps (\emph{dateCreated} and \emph{dateModified})
are in {U}nix time (number of seconds since the Epoch)~\cite{epoch_time} and represented as integers.

\lstset{
    breaklines=true,
    basicstyle=\ttfamily\scriptsize,
    caption={Anonymized {F}ree{BSD} revision in \json\ format.},
    label=code:freebsd-revision-sample
}

\begin{lstlisting}
      {
        "id": 1234567890,
        "type": "DREV",
        "phid": "PHID-DREV-viqvtimavobvsqvgbugp",
        "fields": {
          "title": "Title description",
          "uri": "https://reviews.freebsd.org/D1234567890",
          "authorPHID": "PHID-USER-gnuefszwyfdzrescjihs",
          "status": {
            "value": "published",
            "name": "Closed",
            "closed": true,
            "color.ansi": "cyan"
          },
          "repositoryPHID": "PHID-REPO-tucbfqmbgohbfczzvcfg",
          "diffPHID": "PHID-DIFF-hmkchkgoiochmcmfmgor",
          "summary": "Summary of the code changes.",
          "testPlan": "",
          "isDraft": false,
          "holdAsDraft": false,
          "dateCreated": 3053866284,
          "dateModified": 3211780921,
          "policy": {
            "view": "public",
            "edit": "users"
          }
        },
      ...
\end{lstlisting}

For example, sample \json\ content in~\Cref{code:freebsd-revision-sample}
represents a subset of a record in the \emph{revisions} table.

\section{Conclusions and future work}
\label{sec:conclusions}

We envision several applications for our \phab\ \cor\ dataset.
So far, we have used it to investigate 
\begin{enumerate*}[label=(\alph*),before=\unskip{ }, itemjoin={{, }}, itemjoin*={{, and }}]
    \item the relationship of code changes to acceptance time
    \item the presence of non-productive time during the \cor\ process.
\end{enumerate*}
We are also exploring the dataset for deeper insights into factors
impacting \cor\ acceptance.

The current dataset also opens avenues for new research opportunities.
For example, 
\begin{enumerate*}[label=(\alph*),before=\unskip{ }, itemjoin={{, }}, itemjoin*={{, and }}]
    \item utilizing the formal association between \cor s and bugs
    (tracked by Mozilla project~\cite{bugzilla})
    \item evolution of \diff s by analyzing their subsequent versions
    \item investigating events that occur during \cor\ life cycle
    that other code collaboration tools do not track.
\end{enumerate*}

In this version of the dataset, we chose not to include all the
details about each version of the \diff, such as statistics about
code changes per file, file names, etc.
Our intention here is to avoid duplication of data stored in a source control system.
If such fine-grained data appears to be relevant, the current dataset can be augmented for deeper insights.

\onecolumn

\begin{multicols}{2}
    \bibliographystyle{ACM-Reference-Format}
    \bibliography{phabricator}


\begin{thebibliography}{26}


\ifx \showCODEN    \undefined \def \showCODEN     #1{\unskip}     \fi
\ifx \showDOI      \undefined \def \showDOI       #1{#1}\fi
\ifx \showISBNx    \undefined \def \showISBNx     #1{\unskip}     \fi
\ifx \showISBNxiii \undefined \def \showISBNxiii  #1{\unskip}     \fi
\ifx \showISSN     \undefined \def \showISSN      #1{\unskip}     \fi
\ifx \showLCCN     \undefined \def \showLCCN      #1{\unskip}     \fi
\ifx \shownote     \undefined \def \shownote      #1{#1}          \fi
\ifx \showarticletitle \undefined \def \showarticletitle #1{#1}   \fi
\ifx \showURL      \undefined \def \showURL       {\relax}        \fi
\providecommand\bibfield[2]{#2}
\providecommand\bibinfo[2]{#2}
\providecommand\natexlab[1]{#1}
\providecommand\showeprint[2][]{arXiv:#2}

\bibitem[\protect\citeauthoryear{{B}lender}{{B}lender}{2021}]%
        {blender_phab}
\bibfield{author}{\bibinfo{person}{{B}lender}.}
  \bibinfo{year}{2021}\natexlab{}.
\newblock \bibinfo{title}{{B}lender's {P}habricator instance}.
\newblock
\newblock
\urldef\tempurl%
\url{https://developer.blender.org/}
\showURL{%
Retrieved November 30, 2021 from \tempurl}


\bibitem[\protect\citeauthoryear{Cotet}{Cotet}{2019}]%
        {cotet_2019}
\bibfield{author}{\bibinfo{person}{Dumitru Cotet}.}
  \bibinfo{year}{2019}\natexlab{}.
\newblock \emph{\bibinfo{title}{Crawling Code Review Data From {P}habricator}}.
\newblock \bibinfo{thesistype}{Master's\ thesis}.
  \bibinfo{school}{Friedrich-Alexander University Erlangen-N{\"u}rnberg}.
\newblock
\urldef\tempurl%
\url{https://oss.cs.fau.de/2019/08/07/final-thesis-crawling-code-review-data-from-phabricator/}
\showURL{%
Retrieved November 30, 2021 from \tempurl}


\bibitem[\protect\citeauthoryear{Cotet}{Cotet}{2021}]%
        {phabry_2021}
\bibfield{author}{\bibinfo{person}{Dumitru Cotet}.}
  \bibinfo{year}{2021}\natexlab{}.
\newblock \bibinfo{title}{Phabry}.
\newblock
\newblock
\urldef\tempurl%
\url{https://github.com/dimonco/Phabry}
\showURL{%
Retrieved November 30, 2021 from \tempurl}


\bibitem[\protect\citeauthoryear{c{URL} project}{c{URL} project}{2021}]%
        {curl}
\bibfield{author}{\bibinfo{person}{The c{URL} project}.}
  \bibinfo{year}{2021}\natexlab{}.
\newblock \bibinfo{title}{{C}url}.
\newblock
\newblock
\urldef\tempurl%
\url{https://curl.se/}
\showURL{%
Retrieved November 30, 2021 from \tempurl}


\bibitem[\protect\citeauthoryear{Dickey}{Dickey}{2021}]%
        {diffstat}
\bibfield{author}{\bibinfo{person}{Thomas~E. Dickey}.}
  \bibinfo{year}{2021}\natexlab{}.
\newblock \bibinfo{booktitle}{\emph{diffstat manual}}.
\newblock
\urldef\tempurl%
\url{https://invisible-island.net/diffstat/diffstat.html}
\showURL{%
Retrieved November 30, 2021 from \tempurl}


\bibitem[\protect\citeauthoryear{{FreeBSD}}{{FreeBSD}}{2021}]%
        {freebsd_phab}
\bibfield{author}{\bibinfo{person}{{FreeBSD}}.}
  \bibinfo{year}{2021}\natexlab{}.
\newblock \bibinfo{title}{{FreeBSD}'s {P}habricator instance}.
\newblock
\newblock
\urldef\tempurl%
\url{https://reviews.freebsd.org/}
\showURL{%
Retrieved November 30, 2021 from \tempurl}


\bibitem[\protect\citeauthoryear{Garcia-Molina, Ullman, and
  Widom}{Garcia-Molina et~al\mbox{.}}{2009}]%
        {garcia-molina_database_2009}
\bibfield{author}{\bibinfo{person}{Hector Garcia-Molina},
  \bibinfo{person}{Jeffrey~D. Ullman}, {and} \bibinfo{person}{Jennifer Widom}.}
  \bibinfo{year}{2009}\natexlab{}.
\newblock \bibinfo{booktitle}{\emph{Database Systems: the Complete Book}
  (\bibinfo{edition}{2nd ed} ed.)}.
\newblock \bibinfo{publisher}{Pearson Prentice Hall}, \bibinfo{address}{Upper
  Saddle River, N.J}.
\newblock
\showISBNx{9780131873254}
\newblock
\shownote{OCLC: ocn191926993.}


\bibitem[\protect\citeauthoryear{GitHub}{GitHub}{2016}]%
        {github_code_review_2016}
\bibfield{author}{\bibinfo{person}{GitHub}.} \bibinfo{year}{2016}\natexlab{}.
\newblock \bibinfo{title}{A Whole New Github Universe: Announcing New Tools,
  Forums, and Features}.
\newblock
\newblock
\urldef\tempurl%
\url{https://github.blog/2016-09-14-a-whole-new-github-universe-announcing-new-tools-forums-and-features/}
\showURL{%
Retrieved November 30, 2021 from \tempurl}


\bibitem[\protect\citeauthoryear{Gousios}{Gousios}{2013}]%
        {ghtorrent_dataset}
\bibfield{author}{\bibinfo{person}{Georgios Gousios}.}
  \bibinfo{year}{2013}\natexlab{}.
\newblock \showarticletitle{The {GHTorrent} Dataset and Tool Suite}. In
  \bibinfo{booktitle}{\emph{Proceedings of the 10th Working Conference on
  Mining Software Repositories}} \emph{(\bibinfo{series}{MSR '13})}.
  \bibinfo{publisher}{IEEE Press}, \bibinfo{address}{San Francisco, CA, USA},
  \bibinfo{pages}{233–236}.
\newblock
\showISBNx{9781467329361}


\bibitem[\protect\citeauthoryear{IEEE and Group}{IEEE and Group}{2021}]%
        {epoch_time}
\bibfield{author}{\bibinfo{person}{IEEE} {and} \bibinfo{person}{The~Open
  Group}.} \bibinfo{year}{2021}\natexlab{}.
\newblock \bibinfo{title}{The Open Group Base Specifications Issue 7, 2018
  edition}.
\newblock
\newblock
\urldef\tempurl%
\url{https://pubs.opengroup.org/onlinepubs/9699919799/basedefs/V1_chap04.html#tag_04_16}
\showURL{%
Retrieved November 30, 2021 from \tempurl}


\bibitem[\protect\citeauthoryear{{KDE}}{{KDE}}{2021}]%
        {kde_phab}
\bibfield{author}{\bibinfo{person}{{KDE}}.} \bibinfo{year}{2021}\natexlab{}.
\newblock \bibinfo{title}{{KDE}'s {P}habricator instance}.
\newblock
\newblock
\urldef\tempurl%
\url{https://phabricator.kde.org/}
\showURL{%
Retrieved November 30, 2021 from \tempurl}


\bibitem[\protect\citeauthoryear{{LLVM}}{{LLVM}}{2021}]%
        {llvm_phab}
\bibfield{author}{\bibinfo{person}{{LLVM}}.} \bibinfo{year}{2021}\natexlab{}.
\newblock \bibinfo{title}{{LLVM}'s {P}habricator instance}.
\newblock
\newblock
\urldef\tempurl%
\url{https://reviews.llvm.org/}
\showURL{%
Retrieved November 30, 2021 from \tempurl}


\bibitem[\protect\citeauthoryear{{M}ozilla}{{M}ozilla}{2021a}]%
        {bugzilla}
\bibfield{author}{\bibinfo{person}{{M}ozilla}.}
  \bibinfo{year}{2021}\natexlab{a}.
\newblock \bibinfo{title}{Bugzilla {Main} {Page}}.
\newblock
\newblock
\urldef\tempurl%
\url{https://bugzilla.mozilla.org/home}
\showURL{%
Retrieved November 30, 2021 from \tempurl}


\bibitem[\protect\citeauthoryear{{M}ozilla}{{M}ozilla}{2021b}]%
        {mozilla_phab}
\bibfield{author}{\bibinfo{person}{{M}ozilla}.}
  \bibinfo{year}{2021}\natexlab{b}.
\newblock \bibinfo{title}{{M}ozilla's {P}habricator instance}.
\newblock
\newblock
\urldef\tempurl%
\url{https://phabricator.services.mozilla.com/}
\showURL{%
Retrieved November 30, 2021 from \tempurl}


\bibitem[\protect\citeauthoryear{Mukadam, Bird, and Rigby}{Mukadam
  et~al\mbox{.}}{2013}]%
        {mukadam_2013}
\bibfield{author}{\bibinfo{person}{Murtuza Mukadam}, \bibinfo{person}{Christian
  Bird}, {and} \bibinfo{person}{Peter~C. Rigby}.}
  \bibinfo{year}{2013}\natexlab{}.
\newblock \showarticletitle{Gerrit Software Code Review Data from Android}. In
  \bibinfo{booktitle}{\emph{Proceedings of the 10th Working Conference on
  Mining Software Repositories}} (San Francisco, CA, USA)
  \emph{(\bibinfo{series}{MSR '13})}. \bibinfo{publisher}{IEEE Press},
  \bibinfo{pages}{45--48}.
\newblock
\showISBNx{9781467329361}


\bibitem[\protect\citeauthoryear{Paixao, Krinke, Han, and Harman}{Paixao
  et~al\mbox{.}}{2018}]%
        {crop_dataset}
\bibfield{author}{\bibinfo{person}{Matheus Paixao}, \bibinfo{person}{Jens
  Krinke}, \bibinfo{person}{Donggyun Han}, {and} \bibinfo{person}{Mark
  Harman}.} \bibinfo{year}{2018}\natexlab{}.
\newblock \showarticletitle{{CROP}: Linking Code Reviews to Source Code
  Changes}. In \bibinfo{booktitle}{\emph{2018 IEEE/ACM 15th International
  Conference on Mining Software Repositories (MSR)}}. \bibinfo{pages}{46--49}.
\newblock


\bibitem[\protect\citeauthoryear{Phabricator}{Phabricator}{2021a}]%
        {conduit_tools}
\bibfield{author}{\bibinfo{person}{Phabricator}.}
  \bibinfo{year}{2021}\natexlab{a}.
\newblock \bibinfo{title}{{Community} {Resources}}.
\newblock
\newblock
\urldef\tempurl%
\url{https://secure.phabricator.com/w/community_resources/}
\showURL{%
Retrieved November 30, 2021 from \tempurl}


\bibitem[\protect\citeauthoryear{Phabricator}{Phabricator}{2021b}]%
        {conduit_api_overview}
\bibfield{author}{\bibinfo{person}{Phabricator}.}
  \bibinfo{year}{2021}\natexlab{b}.
\newblock \bibinfo{title}{{Conduit} {API} {Overview}}.
\newblock
\newblock
\urldef\tempurl%
\url{https://secure.phabricator.com/book/phabricator/article/conduit/}
\showURL{%
Retrieved November 30, 2021 from \tempurl}


\bibitem[\protect\citeauthoryear{Phabricator}{Phabricator}{2021c}]%
        {differential_getrawdiff}
\bibfield{author}{\bibinfo{person}{Phabricator}.}
  \bibinfo{year}{2021}\natexlab{c}.
\newblock \bibinfo{title}{differential.getrawdiff}.
\newblock
\newblock
\urldef\tempurl%
\url{https://secure.phabricator.com/conduit/method/differential.getrawdiff/}
\showURL{%
Retrieved November 30, 2021 from \tempurl}


\bibitem[\protect\citeauthoryear{Phabricator}{Phabricator}{2021d}]%
        {differential_query}
\bibfield{author}{\bibinfo{person}{Phabricator}.}
  \bibinfo{year}{2021}\natexlab{d}.
\newblock \bibinfo{title}{differential.query}.
\newblock
\newblock
\urldef\tempurl%
\url{https://secure.phabricator.com/conduit/method/differential.query/}
\showURL{%
Retrieved November 30, 2021 from \tempurl}


\bibitem[\protect\citeauthoryear{Phabricator}{Phabricator}{2021e}]%
        {phabricator_usage}
\bibfield{author}{\bibinfo{person}{Phabricator}.}
  \bibinfo{year}{2021}\natexlab{e}.
\newblock \bibinfo{title}{Organizations Using Phabricator}.
\newblock
\newblock
\urldef\tempurl%
\url{https://secure.phabricator.com/w/usage/companies/}
\showURL{%
Retrieved November 30, 2021 from \tempurl}


\bibitem[\protect\citeauthoryear{Phabricator}{Phabricator}{2021f}]%
        {phabricator_api}
\bibfield{author}{\bibinfo{person}{Phabricator}.}
  \bibinfo{year}{2021}\natexlab{f}.
\newblock \bibinfo{title}{{Query}: {Modern} {Methods}}.
\newblock
\newblock
\urldef\tempurl%
\url{https://secure.phabricator.com/conduit/}
\showURL{%
Retrieved November 30, 2021 from \tempurl}


\bibitem[\protect\citeauthoryear{Phabricator}{Phabricator}{2021g}]%
        {transaction_search}
\bibfield{author}{\bibinfo{person}{Phabricator}.}
  \bibinfo{year}{2021}\natexlab{g}.
\newblock \bibinfo{title}{transaction.search}.
\newblock
\newblock
\urldef\tempurl%
\url{https://secure.phabricator.com/conduit/method/transaction.search/}
\showURL{%
Retrieved November 30, 2021 from \tempurl}


\bibitem[\protect\citeauthoryear{Phacility}{Phacility}{2021}]%
        {phacility}
\bibfield{author}{\bibinfo{person}{Phacility}.}
  \bibinfo{year}{2021}\natexlab{}.
\newblock \bibinfo{title}{Phacility - {Home}}.
\newblock
\newblock
\urldef\tempurl%
\url{https://www.phacility.com/}
\showURL{%
Retrieved November 30, 2021 from \tempurl}


\bibitem[\protect\citeauthoryear{Tsotsis}{Tsotsis}{2011}]%
        {tsotsis_meet_phabrictor}
\bibfield{author}{\bibinfo{person}{Alexia Tsotsis}.}
  \bibinfo{year}{2011}\natexlab{}.
\newblock \bibinfo{title}{Meet {Phabricator}, {The} {Witty} {Code} {Review}
  {Tool} {Built} {Inside} {Facebook}}.
\newblock
\newblock
\urldef\tempurl%
\url{https://social.techcrunch.com/2011/08/07/oh-what-noble-scribe-hath-penned-these-words/}
\showURL{%
Retrieved November 30, 2021 from \tempurl}


\bibitem[\protect\citeauthoryear{Yang, Kula, Yoshida, and Iida}{Yang
  et~al\mbox{.}}{2016}]%
        {gerrit_dataset}
\bibfield{author}{\bibinfo{person}{Xin Yang}, \bibinfo{person}{Raula~Gaikovina
  Kula}, \bibinfo{person}{Norihiro Yoshida}, {and} \bibinfo{person}{Hajimu
  Iida}.} \bibinfo{year}{2016}\natexlab{}.
\newblock \showarticletitle{Mining the Modern Code Review Repositories: A
  Dataset of People, Process and Product}. In
  \bibinfo{booktitle}{\emph{Proceedings of the 13th International Conference on
  Mining Software Repositories}} (Austin, Texas) \emph{(\bibinfo{series}{MSR
  '16})}. \bibinfo{publisher}{Association for Computing Machinery},
  \bibinfo{address}{New York, NY, USA}, \bibinfo{pages}{460–463}.
\newblock
\showISBNx{9781450341868}
\urldef\tempurl%
\url{https://doi.org/10.1145/2901739.2903504}
\showDOI{\tempurl}


\end{thebibliography}
\end{multicols}

\end{document}